# EQUILIBRIUM OF SPHERES WITH LOCAL ANISOTROPY IN POSTNEWTONIAN GRAVITY. APPLICATION TO WHITE DWARFS

E. S. Corchero

Departamento de Física. Universidad de Cantabria. Santander

**Abstract**

Static solutions of white dwarfs with spherical symmetry and local anisotropy are studied in the post-Newtonian approximation. It is argued that the condition for equilibrium must be that the total energy is a minimum for given baryon number and the question whether there is local isotropy or anisotropy inside the star should follow from that condition, rather than be postulated "a priori". It is shown show that, in post-Newtonian gravity, there are stable configurations with local anisotropy for masses above the Chandrasekhar limit.

Address

E. Santos Corchero

Facultad de Ciencias

Avenida de los Castros s/n

39005 Santander. Spain

FAX nº 34-42-201402

Email <santose besaya.unican.es>

## 1. INTRODUCTION

It is well known that both Newtonian gravity and general relativity allow spherically symmetric solutions with local anisotropy and that anisotropy may have an important effect on the stability [1]. Einstein himself considered in 1939 a model with local anisotropy, consisting of a spherical cluster of particles each moving in a circle, and concluded that "the essential result of this investigation is a clear understanding as to why the `Scwartzchild singularities´ do not exist in physical reality" [2]. (It is ironic that the same year Oppenheimer and Snyder [3] published the first calculation demostrating the formation of a black hole, and Einstein´s paper was cast into oblivion).

In more recent times, local isotropy has been the common assumption in studies of stellar evolution. Nevertheless, different scenarios have been proposed which incorporate local anisotropy into the modelling of stellar objects, e.g. pion condensation, phase transitions, boson stars and crystalization in white dwarfs, among others (see [1] and references therein). A standard assumption in these approaches is that the departure from local isotropy is caused by the equation of state of the matter forming the star being essentially anisotropic. In this paper we shall make a different assumption, namely that anisotropy may be produced by a dynamical effect of the gravitational field itself even if the equation of state of the fluid forming the star is isotropic in free space.

We shall argue that the condition of equilibrium of white dwarfs is that the total energy of the star is a minimum for given number of baryons. We shall show that local anisotropy is a straightforward consequence of this general condition in the case of massive white dwarfs. We consider only Newtonian and post-Newtonian approximations, leaving for the future the study of stars where the gravitational field must be treated using general relativity.

## 2. NEWTONIAN WHITE DWARFS WITH LOCAL ANISOTROPY

A white dwarf consists of a plasma of protons, heavier nuclei (mainly helium) and electrons. We shall assume that the chemical composition of the plasma is homogeneous throughout the star. The temperature of the plasma inside the star may be taken as zero Kelvin (this is because the quantum zeropoint energy of the electrons is much greater than the thermal energy of the plasma, see e.g. Ref.4). As a consequence the star is, to a good approximation, an isolated physical system at zero temperature and the standard condition for stable equilibrium holds, namely the total energy should be a minimum for given baryon number (or, in Newtonian gravity, minimum energy for a given mass, because mass is proportional to baryon number in this case). *We should not impose additional constraints.* In particular we must not assume "a priori" that the pressure is isotropic at every point inside the star, but the general condition of equilibrium (minimal energy) is what will determine whether the pressure is isotropic or not.

In Newtonian gravity the statement of the problem is straightforward. We must find the ground state of a quantum system with Hamiltonian

$$H = \Sigma_j \sqrt{(m_j^2 c^4 + \mathbf{p}_j^2 c^2)} + \Sigma_{ij} (Z_i Z_j e^2 - G m_i m_j) r_{ij}^{-1}, \qquad (2.1)$$

where i, j label two arbitrary constituent particles (nuclei and/or electrons). $Z_j e$ ($m_j$, $\mathbf{p}_j$) is the charge (mass, momentum) of particle j, and G (c) is Newton´s constant (the velocity of light). In the Hamiltonian (2.1) we have ignored magnetic interactions, electron-positron pair creation and the possibility of neutronization (the reaction of an electron and a proton to give a neutron and a neutrino). To a good approximation magnetic interactions and pair creation may be neglected. In contrast, neutronization may be relevant in some cases [4] but we shall ignore it for the sake of simplicity ( we shall comment on it in section 7).

The ground state of the Hamiltonian (2.1) cannot be found without making some approximations. An obvious one derives from the weakness and long range character of the gravitational field (the electrostatic interaction is effectively short ranged because atractions and repulsions cancel each other at a long distance). Hence it follows that the typical distance where the gravitational field changes inside the star is many orders of magnitude greater than the de Broglie wavelength of the constituent particles. This allows us to rewrite the Hamiltonian (2.1) approximating the last term in the form

$$H = \Sigma_j \sqrt{(m_j^2 c^4 + \mathbf{p}_j^2 c^2)} + \Sigma_{ij} Z_i Z_j e^2 r_{ij}^{-1} + \Sigma_j m_j \Phi(r_j),$$

$$\Phi(r_j) = -G \langle \Psi | \Sigma_i m_i r_{ij}^{-1} | \Psi \rangle, \qquad (2.2)$$

where $\Phi(r_j)$ is the gravitational potential at $r_j$ and $\Psi$ the wavefunction of the plasma. Our problem is now to find the ground state of a plasma in a given external potential $\Phi$, but constrained by the consistency requirement that $\Phi$ is the potential created by the plasma itself. So stated the problem is still formidable because it involves the full machinery of quantum many body theory, but there is a good enough approximation which is described in the following.

We may consider that the particles are independent, except for some averaged electrostatic interaction. As the velocity of nuclei is much smaller than that of the electrons, we may treat them as being at rest. Furthermore, we may replace the nuclei by a jellium of positive charge so that the net charge is zero, which is a standard approximation for a plasma at zero temperature. After that the plasma becomes a degenerate electron gas immersed in a background of positive charge. Such a gas is fully characterized by the size and form of the Fermi surface at every point of space. In fact, from it we may obtain all functions of position needed to solve the problem of minimizing the energy insuring gravitational equilibrium (i.e. the fulfillment of the second eq.(2.2) ). These functions are: the electron density, n, the energy density, u, and the stress tensor, $p_{ij}$, which we get as follows (we shall use Planck units $G = c = h = 1$ from now on):

$$n = (4\pi^3)^{-1} \int d^3k, \qquad nu = (4\pi^3)^{-1} \int [\sqrt{(m_e^2 + k^2)} - m_e] \, d^3k,$$

$$p_{ij} = (4\pi^3)^{-1} \int k_i k_j [\sqrt{(m_e^2 + k^2)}]^{-1} d^3k, \qquad (2.3)$$

where the integrals should be extended to the interior of the Fermi surface. As a consequence of the spherical symmetry, which we will assume for the (nonrotating) star, there is one privileged direction at every point inside the star, the radial direction. Thus the Fermi surface at every point will possess rotational symmetry around the radial direction. This implies that there are only two principal stresses, namely the radial pressure, $p_r$, and the transverse pressure, $p_t$, for which the last eqs.(2.3) become eqs.(3.9) and (3.10) (see below).

Now the problem of finding the ground state of the Hamiltonian (2.1) has been reduced to the problem of finding the size and form of the Fermi surface at every point inside the star in such a way that: 1) gravitational (hysrostatic) equilibrium exists at any point, and 2) the total energy of the star is a minimum for given total baryon number. In Newtonian gravity the hydrostatic condition for equilibrium is (we *do not* assume isotropic pressure although we do not exclude this possibility)

$$dp_r / dr + 2 ( p_r - p_t ) / r = - m \rho_0 / r^2, \qquad m(r) = 4 \pi \int_o^r \rho_0 r^2 dr, \qquad (2.4)$$

where we have introduced the mass density, $\rho_0$, defined by

$$\rho_0(r) = \mu \, m_B \, n(r) . \qquad (2.5)$$

$m_B$ being the mean baryon rest mass and $\mu$ the mean number of baryons per electron. (For the derivation of eq.(2.4) see eq.(5.3) below). The mass and energy of the star are given by

$$M = \int \rho_0 \, d^3r , \qquad E = \int \{ u - m / r \} \, \rho_0 \, d^3r . \qquad (2.6)$$

In the present paper we shall consider only the case of small anisotropy, that is small deformation of the Fermi surface with respect to a sphere (the calculation in the general case is more involved and will be made elsewhere). Thus the Fermi surface may be characterized by just two functions of the radius, and we may write the mass density, $\rho_0$, the energy density, u, the radial pressure, $p_r$. and the transverse pressure, $p_t$, in terms of those two functions using eqs.(2.2) and (2.3) as is explained in the next section. After that we shall be able to find the functions $\rho_0(r)$, $u(r)$, $p_r(r)$ and $p_t(r)$, fulfilling the hydrostatic equilibrium eq.(2.4) and making E a minimum for given M, eqs.(2.6). The practical procedure to do that is described in section 4.

It is interesting to point out that eq.(2.4) would become the conventional equation of hydrostatic equilibrium if we were assuming isotropic pressure, that is

$$dp/dr = -m\rho_0/r^2, \qquad p = p_t = p_r. \qquad (2.7)$$

In this case both requirements, the energy E ( eq.(2.6)) being a minimum and the hydrostatic equilibrium (eq.(2.7) ) are equivalent conditions. However, if we do not assume local isotropy "a priori", the hydrostatic equilibrium ( now given by eq.(2.4)) and the minimal energy are independent conditions and both should hold.

## 3. ANISOTROPIC DEGENERATE ELECTRON GAS.

As said above, we shall take the electrons as noninteracting particles, although corrections to this simple model are well known. In a degenerate electron gas, anisotropy appears if the Fermi sphere of the electrons is deformed. If the deformation is small we may write

$$k_F(\theta) = k_{F0}[1 - y P_2(z)], \quad z = \cos\theta, \qquad (3.1)$$

where $\theta$ is the angle with respect to a given direction (the radial direction in the case of a star), $P_2(z)$ is a Legendre polynomial and y is a parameter measuring the fractional anisotropy. In (3.1) a dipole term is excluded in order that there is no net electric current at any point, but higher multipoles of even order should be introduced if the anisotropy were not small.

From (3.1) it is easy to get the number density of electrons (compare with eq.(2.3) )

$$n = (2\pi^2)^{-1} \int_{-1}^{1} dz \int_0^{kF} k^2 \, dk = (3\pi^2)^{-1} k_{F0}^3 \, (1+3/5 \; y^2+2/35 \; y^3).$$
(3.2)

The energy per electron, u, or the energy density, nu, may be also obtained easily. We have

$$n\, u = (2\pi^2)^{-1} \int_{-1}^{1} dz \int_0^{kF} k^2 \, dk \, [\sqrt{(m_e^2 + k^2)} - m_e \,] =$$

$$= (2\pi^2)^{-1} m_e^4 \int_{-1}^{1} dz \int_0^{1-yP(z)} s^2 \, [\sqrt{(1+ v^2 s^2)} - 1 \,] \, ds,$$
(3.3)

where $m_e$ is the electron mass and we have introduced the new variables

$$v = k_{F0} / m_e , \qquad\qquad s = k / k_{F0} .$$
(3.4)

The calculation of (3.3) is straightforward using the expansion

$$\int_0^{1-\varepsilon} f(s) \, ds = \int_0^1 f(s) \, ds \; - \; \varepsilon \, f(1) + \tfrac{1}{2} \, \varepsilon^2 \, (df/ds)_{s=1} + O(\varepsilon^3).$$
(3.5)

We get

$$n\, u = \pi^{-2} \, m_e^4 \, v^3 \, (\chi + \phi \, y^2) + O(y^3) , \quad u = 3 \, m_e \, [\, \chi + (\phi - 3/5 \, \chi) \, y^2 \,] + O(y^3),$$
(3.6)

where we have introduced the functions

$$\chi(v)=\int_0^1 s^2 \, [\sqrt{(1+ v^2 s^2)} -1] ds = (8v^3)^{-1} \{ v \, (1+ 2v^2) \, \sqrt{(1+v^2)} \; - \; \log[v+\sqrt{(1+v^2)}]\},$$
(3.7)

$\phi(v) = 1/10 \; (d/ds[s^2 \; (\sqrt{(1+v^2 s^2)} \; -1)] \;)_{s=1} = 1/10 \; (2+3v^2) \; /\sqrt{(1+v^2)} \quad - 1/5.$
(3.8)

Similarly, we may obtain the radial and transverse pressures

$$p_r = (2\pi^2)^{-1} \int_{-1}^{1} z^2 \, dz \int_{0}^{kF} k^4 \, dk \; [\sqrt{(m_e^2 + k^2)}]^{-1} =$$

$$= (2\pi^2 h^3)^{-1} m_e^4 v^5 \int_{-1}^{1} z^2 \, dz \int_{0}^{1-yP(z)} s^4 \; [\sqrt{(1+ v^2 s^2)}]^{-1} \, ds =$$

$$= (3\pi^2)^{-1} \; m_e^4 \; v^4 \; ( \chi' \; - \; 2/5 \; [v/\sqrt{(1+ v^2)}] \; y \; + \; 11/7 \; \phi' \; y^2 \; ) \; + \; O(y^3) \; ,$$
(3.9)

$$p_t = (2\pi^2)^{-1} \int_{-1}^{1} \tfrac{1}{2} (1- z^2) \, dz \int_{0}^{kF} k^4 \, dk \; [\sqrt{(m_e^2 + k^2)}]^{-1} =$$

$$= (3\pi^2)^{-1} \; m_e^4 \; v^4 \; ( \chi' \; + \; 1/5 \; [v/\sqrt{(1+ v^2)}] \; y \; + \; 5/7 \; \phi' \; y^2 \; ) \; + \; O(y^3) \; ,$$
(3.10)

where $\chi' = d\chi/dv$ and $\phi' = d\phi/dv$. The mean pressure is

$$p = 1/3 \; (p_r + 2 \; p_t ) = (3\pi^2)^{-1} \; m_e^4 \; v^4 \; ( \chi' + \phi' \; y^2 ) + O(y^3) . \qquad (3.11)$$

It is not difficult to check that p is related to the electron density, n, and the energy per electron, u, by a standard thermodynamic relation which, in the case of anisotropy, should be written

$$p = n^2 \; \partial u/\partial n|_{y=const} = n^2 \; (\partial u/\partial v)_y \; /(\partial n/\partial v)_y \; . \qquad (3.12)$$

It is convenient to replace the anisotropy parameter y by another one x:

$$x = (p_t - p_r)/3 \; p = \; 1/5 \; [v/\sqrt{(1+ v^2)}] \; \chi'^{-1} \; y \; - \; 2/7 \; \phi' \; \chi'^{-1} \; y^2 \; + \; O(y^3) . \qquad (3.13)$$

From the definition it follows that $x \in [-1, 1/2]$. If we eliminate $v$, $k_{F0}$ and $y$ amongst eqs.(3.2), (3.6) and (3.13) we may obtain the energy per electron, $u$, in terms of the electron density, $n$, and the anisotropy parameter, $x$:

$$u = 3\, m_e \{ \chi(v) + [25\, \phi(v) - 15\, \chi(v)]\, (1+v^2)\, v^{-2}\, \chi(v)'^{\,2}\, x^2 \} + O(x^3), \quad (3.14)$$

where $v$ stands here for $(3\pi^2 n)^{1/3}/m_e$. In particular, in the nonrelativistic ($v \ll 1$) and ultrarelativistic ($v \gg 1$) regimes we have

$$u = 3/10\, (3\pi^2 n)^{2/3}/m_e\, (1+x^2) \text{ if } v \ll 1, \quad u = 3/4\, (3\pi^2 n)^{1/3}\, (1 + 5/8\, x^2) \text{ if } v \gg 1. \quad (3.15)$$

Hence, the mean pressure, $p$, may be obtained from (3.12) and, taking (3.13) into account, the radial and transverse pressures are given by

$$p_r = (1 - 2x)\, p, \quad p_t = (1 + x)\, p. \quad (3.16)$$

Also we shall redefine $u$ as the energy per unit mass dividing the previous $u$ by $\mu m_B$. We are interested in white dwarfs of high mass (close to the Chandrasekhar limit) where the electrons are ultrarelativistic. But the approximation of the second eq.(3.15) is not enough and we require corrections to that expression up to $O(v^{-2})$. Nevertheless we shall consider only cases where the local anisotropy is small ($x \ll 1$) and this allows us to introduce corrections for anisotropy only in the leading term of the equation of state. Consequently we shall write

$$u = u_0 + u_1 + u_A,\quad u_0 = 3\, K\, \rho_0^{1/3},\quad u_1 = K' + 3\, K''\rho_0^{-1/3},\quad u_A = 15/8\, K\, \rho_0^{1/3}\, x^2, \quad (3.17)$$

where the constants $K$, $K'$ and $K''$ may be obtained from (3.14) and (2.5). Their values are well known (see e.g. Ref. 4) and only the numerical coefficient of the anisotropy correction $u_A$ is a new result. In our work of the following sections both $u_1$ and $u_A$ will be considered corrections of the same order to the

main term $u_0$. The mean pressure may be obtained from (3.17) and the thermodynamic relation (3.12). We get

$$p = p_0 + p_1 + p_A, \quad p_0 = K \rho_0^{4/3}, \quad p_1 = -K'' \rho_0^{2/3}, \quad p_A = 15/8 \, K \rho_0^{4/3} x^2.$$
(3.18)

In a homogeneous gas without external forces, the positivity of the anisotropy correction (term with $x^2$ in eq.(3.17)) implies that the minimal energy density for given density of particles corresponds to $x=0$, that is local isotropy. But we claim that the condition of minimal energy is the only cause why the proton-electron plasma at zero temperature is isotropic in free space. As is very well known the Fermi surface of the electron gas may be deformed by electrostatic fields (e.g. in metals) and the gas pressure is no longer isotropic. We think that there is no reason to *postulate* that it cannot be deformed also by the gravitational field, a postulate nevertheless made in conventional teatments of relativistic stars.

## *4. NEWTONIAN EQUILIBRIUM WITH LOCAL ANISOTROPY*

After the results of the two previous sections, the theory of Newtonian white dwarfs is straightforward. (If the mass is low enough the star behaves like a polytrope with $\gamma = 5/3$, the equilibrium state corresponding to isotropic pressure [5], but here we shall be concerned with the more interesting case of high mass stars, whose equation of state is given by (3.18).) In the conventional treatment [4] it is *postulated* that $p_r = p_t$, so that the problem reduces to solving eq.(2.7) (see the last paragraph of section 2). In practice a variational method is used in which a one-parameter family of configurations is selected with the star approximated by a polytrope. That is, we change variables:

$$\rho_0(r) = \rho_c \, \theta^3, \qquad r = [\, 4\pi \, \xi_1^2 \, |\theta'(\xi_1)| \,]^{-1/3} \, M^{1/3} \, \rho_c^{-1/3} \, \xi,$$
(4.1)

and assume that $\theta(\xi)$ is the solution of the Lane-Emden equation of index n=3. The central density $\rho_c$ is the parameter of the family of configurations. Thus eq.(2.6) gives the energy of the star in terms of standard integrals and we obtain

$$E_I = ( A M - B M^{5/3} ) \rho_c^{1/3} - K' M + C M \rho_c^{-1/3}, \qquad (4.2)$$

where A, B, C and K' are well known positive constants [4] and the subindex of $E_I$ stands for "isotropic". The condition of stationarity of the energy for given mass (i.e., $dE/d\rho_c = 0$) leads to the equilibrium density

$$\rho_c = C^{3/2} ( A - B M^{2/3} )^{-3/2}, \qquad (4.3)$$

and stable equilibrium (i.e., $d^2E/d\rho_c^2 > 0$) exists if and only if the mass is less than the Chandrasekhar limit

$$M < M_{Ch} \cong ( A / B )^{3/2}. \qquad (4.4)$$

Untill here the conventional theory, involving the postulate of isotropic pressure (i.e. $x(r) = 0$). If we allow for local anisotropy we shall search for pairs of functions, $\rho(r)$ and $x(r)$, fulfilling (2.4) and (3.16), and select amongst them the pair which makes the energy a minimum. An exact solution of this problem would be quite involved and we shall use a variational (approximate) method which we describe in the following.

We begin by analyzing more closely the hydrostatic equilibrium eq. (2.4). Taking (3.16) into account it may be written

$$d [p (1-2 x)] /dr = - m \rho_0 /r^2 + 6 p x /r . \qquad (4.5)$$

A formal integration of (4.5) gives

$$x = ( 2 r^3 p )^{-1} \int_o^r r^3 dr [ p' + m \rho_0 / r^2 ] . \qquad (4.6)$$

This is an exact equation (within Newtonian gravity) but it is not closed. In fact, it is an integro-differential equation because p depends on x (see (3.18)). Fixed the function $\rho_0(r)$, eq.(4.6) gives the unique solution of (4.5) which is regular at the origin ( it begins with $x = a\, r^2 + b\, r^4 + ...$ ). We also need that x is regular at the star surface, where the pressure vanishes. An obvious necessary condition is that the integral of the right hand side of (4.6) goes to zero when r→R, R being the star radius. This gives, after an integration by parts,

$$\int_0^R ( m\, \rho_0 /r )\, d^3r = 3 \int_0^R p\, d^3r \Rightarrow$$

$$3 \int_0^R p_A\, d^3r = \int_0^R ( m\, \rho_0 /r )\, d^3r - 3 \int_0^R p_0\, d^3r - 3 \int_0^R p_1\, d^3r \geq 0 . \qquad (4.7)$$

This relation may be called the virial theorem with local anisotropy, and becomes the standard virial theorem when $x = 0$. The inequality in (4.7) follows from the nonegativity of $p_A$ and it is a necessary condition for the function x(r), obtained by solving (4.6), being physical (i.e. $x \in [-1, 1/2]$ everywhere).

Now we select a family of locally anisotropic star configurations, using the change of variables (4.1) and assuming that $\theta(\xi)$ is the solution of the Lane-Emden equation of index n=3. Thus for every value of $\rho_C$ we obtain a function $\rho_0(r)$ and, hence, we get x(r) by solving eq.(4.6). For any pair $\{\rho_0(r), x(r)\}$ the energy of the star becomes

$$E = \int \{ u_o(r) + u_1(r) + u_A(r,x) - m / r \}\, \rho_0\, d^3r . \qquad (4.8)$$

Performing the integrals we may calculate the energy in terms of the central density. The (approximate) equilibrium configuration of the star will be the one with minimal energy within the family. All the integrals in (4.8) are straightforward except the third one, involving x(r), which cannot be evaluated without finding explicitly the solution of eq.(4.6). We get

$$E = ( A M - B M^{5/3} )\, \rho_c^{1/3} - K' M + C M\, \rho_c^{-1/3} + \int u_A(r,x)\, \rho(r)\, d^3r . \qquad (4.9)$$

We may avoid the calculation of the last integral using the virial theorem (4.7). In fact, performing the integrals in (4.7), except the one involving $p_A$, we get

$$( B\ M^{5/3} - A\ M )\ \rho_c^{1/3} + C\ M\ \rho_c^{-1/3} = 3 \int p_A(r,x)\ d^3r \geq 0. \quad (4.10)$$

On the other hand, a comparison of (3.17) and (3.18) shows that $u_A\ \rho = p_A$, which allows us to get, from (4.9) and (4.10)

$$E = - K'\ M + 2\ C\ M\ \rho_c^{-1/3} \text{ if } E \geq E_I\ ,\quad E \cong E_I \text{ otherwise}, \quad (4.11)$$

where $E_I$ is given by (4.2). For any mass above the Chandrasekhar limit (4.4), the inequality (4.10) (i. e. $E \geq E_I$) is obviously fulfilled (remember that A, B and C are positive), but the energy (4.11) has no minimum for finite $\rho_c$, it always decreases with increasing central density. This means that there are no equilibrium configurations with anisotropic pressure and the star is unstable against collapse. However, if the mass is below the limit (4.4), the minimum of (4.11) compatible with (4.10) corresponds to the density (4.3) and the equality sign $E = E_I$, that is local isotropy. We see that our method agrees with the conventional one in this case, and we predict stable equilibrium with $\rho_c$ given by (4.3) below the Chandrasekhar limit (4.4).

We must point out that both our calculation (leading to (4.11)) and the conventional one (leading to (4.2)) rest upon variational approximations, i.e. minimizing the energy only within a restricted family of configurations. But our accuracy is superior because we select a family of *exact* solutions of the hydrostatic equilibrium eq.(2.4), whilst the conventional treatment selects a family of *approximate* solutions. Consequently, above the Chandrasekhar limit our treatment is more correct. But below that limit, where we do not find equilibrium configurations with local anisotropy, the conventional treatment provides probably a good approximation. Although rigorous results (for some results see Ref.5) and/or more exact (numerical) calculations would be wellcome, we may conclude that allowing for local anisotropy in Newtonian white dwarfs gives essentially no new result. For any mass below the limit (4.4) the equilibrium configuration is locally isotropic, and there are no equilibrium configurations with mass above that limit. Newtonian gravity predicts the collapse of white dwarfs with mass above the Chandrasekhar limit. In the next two sections we show that the conclusion changes dramatically in post-Newtonian gravity.

## 5. LOCAL ANISOTROPY IN POST-NEWTONIAN APPROXIMATION

Although the hydrostatic equilibrium equation with local anisotropy is well known [1,6], we shall rederive it for the sake of clarity. Solutions of the Einstein equations with spherical symmetry may be most easily obtained using curvature coordinates. The metric is

$$ds^2 = e^\alpha \, dr^2 + r^2 \, (d\theta^2 + \sin^2\theta \, d\phi^2) - e^\beta \, dt^2. \quad (5.1)$$

Here we shall consider only static fields, where $\alpha$ and $\beta$ depend on the single coordinate r. Using the notation $\alpha'$ for $d\alpha/dr$ and $\alpha''$ for $d^2\alpha/dr^2$, and similar for $\beta$, Einstein's equations may be written

$$- 8\pi \, p_r = G_{11} = r^{-2} \, [1 - e^{-\alpha} \, (1 + r \, \beta')],$$

$$- 8\pi \, p_t = G_{22} = G_{33} = 1/4 \, e^{-\alpha} \, [-2 \, \beta'' - \beta'^2 + \alpha' \, \beta' + 2 \, r^{-1} \, (\alpha' - \beta')],$$

$$8\pi \, \rho = G_{44} = r^{-2} \, [1 - e^{-\alpha} \, (1 - r \, \alpha')]. \quad (5.2)$$

Obtaining the condition of hydrostatic equilibrium from eq.(5.2) is straightforward. The third equation allows getting $\alpha$ in terms of the density $\rho$. Hence the first equation gives $\beta$ in terms of $\rho$ and $p_r$. If these expressions for $\alpha$ and $\beta$ are put into the second eq.(5.2) we obtain

$$dp_r/dr + 2 \, (p_r - p_t) \, /r = - m \, \rho/r^2 \, [1 + p_r/\rho] \, [1 + 4\pi \, r^3 \, p_r/m] \, [1 - 2 \, m \, /r]^{-1},$$

$$m(r) = 4\pi \int_0^r \rho \, r^2 \, dr. \tag{5.3}$$

Obviously this equation becomes the conventional equation of hydrostatic equilibrium if we assume local isotropy ($p_t = p_r = p$).

Einstein´s eqs.(5.2) do not imply local isotropy, but only a relation between anisotropy of pressure and anisotropy of the curvature of space, namely

$$8\pi \, (p_t - p_r) = R_{22} - R_{11}, \qquad R_{33} = R_{22}, \tag{5.4}$$

where $R_{\mu\nu}$ are components of the Ricci tensor, the last equality being a consequence of the spherical symmetry. The connection between anisotropy of matter pressure and anisotropy of space curvature leads to the conjecture that local anisotropy may be more natural and relevant in general relativity than in Newtonian gravity. Another argument is that the effective gravitational force depends only on position in Newtonian theory, but depends on *the velocity vector* in general relativity (compare the right hand sides of (2.4) and (5.3), taking into account the relation (2.3) between stress tensor and linear momentum). This conjecture is supported by the calculations reported in the present article.

As is well known, in order to solve the hydrostatic equilibrium eq.(5.3) we need, in addition to boundary conditions, a relation between $\rho$, $p_r$ and $p_t$. This should be obtained from the equation of state which, in general relativity, is usually written in the form of a relation between the energy (or mass) density, $\rho(r)$, and the number density of baryons. In the study of white dwarfs in post-Newtonian (PN) approximation it is more convenient to use the mass density $\rho_0$ (2.5) rather than the baryon density. The total energy (or mass) density will be

$$\rho = \rho_0 \, (1 + u), \tag{5.5}$$

where u is given by (3.17).

In PN approximation the hydrostatic equilibrium may be obtained from eq.(5.3) by neglecting terms of second order in p/ρ and m/r. Using (3.16) we get

$$d[p(1-2x)]/dr = -m\rho/r^2 - mp(1-2x)/r^2 - 4\pi r p(1-2x) - 2 m^2\rho/r^3 + 6p\, x/r. \quad (5.6)$$

Similarly as we obtained eq.(4.6) from (4.5), a formal integration of (5.6) gives

$$x(r) = (2r^3 p)^{-1} e^{-I(r)} \int_0^r e^{I(r')} r'^3 dr' [dp/dr' + m\rho/r'^2 + m p/r'^2 + 4\pi r' \rho p + 2 m^2\rho/r'^3]$$

$$\cong (2r^3 p)^{-1} \int_0^r r'^2 dr' \{[1+I(r')-I(r)][m\rho/r' - 3p] + 2 m^2\rho/r'^2\}, \quad I(r) = \int_0^r r'^{-2} d(mr'), \quad (5.7)$$

where, in the final expression for x(r), we have neglected terms of second order, to be consistent with the PN approximation, and performed several integrations by parts. The condition that x is finite at the star surface leads, after an integration by parts, to the virial theorem in the PN approximation with local anisotropy:

$$\int d^3r \left\{ [1 - \int_r^R r'^{-2} d(mr')][m\rho/r - 3p] + 2 m^2 \rho_o/r^2 \right\} = 0. \quad (5.8)$$

where we have retained only terms up to first order in p/ρ and m/r, and substituted $\rho_o$ for ρ in the last term as is consistent with the PN approximation.

The problem of equilibrium closely parallels the Newtonian one and may be solved as follows. Taking into account that the radius R of a star is defined by the condition p(R) = 0, and therefore $p_r(R) = p_t(R) = 0$, the total mass $M_{tot}$ and the number of baryons N are given by

$$M_{tot} = 4\pi \int_0^R \rho r^2 dr, \quad N = 4\pi \int_0^R n [1 - 2m/r]^{-1/2} r^2 dr. \quad (5.9)$$

In PN approximation it is convenient to define a mass M that includes only the rest mass of nuclei and electrons, in the form

$$M= 4\pi \int_o^R [1-2m/r]^{-1/2}\rho_0\, r^2 dr. \qquad (5.10)$$

The energy of the star is defined as the difference $(M_{tot} - M) c^2$. We get (c = 1)

$$E = 4\pi \int_o^R \rho_o\, r^2\, dr\, (u - m/r - 3/2\, m^2/r^2), \qquad (5.11)$$

where, as usual, we have neglected terms of second order. The condition of equilibrium with local anisotropy is that the energy E (5.11) is a minimum for fixed M (5.10) amongst configurations fulfilling the hydrostatic equilibrium eq. (5.6). The practical procedure is described in the next section.

## 6. WHITE DWARFS WITH LOCAL ANISOTROPY IN PN THEORY

We summarize the conventional theory (i.e. postulating local isotropy from the very beginning) of white dwarfs in the PN approximation [4]. We should evaluate (5.11), using the expression for u given in (3.17) with x=0, perform the change of variables (4.1) and assume that $\theta(\xi)$ is the Lane-Emden function of index n=3. But there is a difficulty because it is the mass M, given by (5.10), rather than the volume integral of $\rho_0$ (4.6) which should be fixed in the variational approach. A method to solve the problem is to change from the radial coordinate r to the new one r´ related to the first by

$$4\pi\, r'^2\, dr' = dV = 4\pi\, [1- 2\, m/r]^{-1/2} r^2\, dr, \qquad (6.1)$$

dV being the invariant volume element. Then it is straightforward, although lengthy [4], to get

$$E_I = (AM - BM^{5/3})\, \rho_c^{1/3} - K´M + C\, M\, \rho_c^{-1/3} - [D_1 M^{7/3} + D_2(A/B)\, M^{5/3}]\, \rho_c^{2/3}. \qquad (6.2)$$

The values of the constants are given in Ref.4, except $D_1=0.02543$, $D_2=0.89285$. The condition that (6.2), as a function of $\rho_c$, has a minimum implies that stable equilibrium is possible only if the mass is below the limit

$$M < (A/B)^{3/2} [1 - 2 (A D^2 C B^{-4})^{1/3}], \quad D = D_1+ D_2, \quad (6.3)$$

which is very close to the Chandrasekhar limit (4.4).

We pass to the theory with allowance for local anisotropy. We should select $\rho_0(r)$ as in (4.1) and get $x(r)$ from (5.7). The energy is obtained adding to $E_I$ (6.2) the integral of the last term of (2.17), but we may simplify a lot the calculation by subtracting (5.8) from (5.11). This gives, after some algebra,

$$E = \int d^3r \{\rho_o u_1 - 3p_1 + 3K\rho_0^{4/3} m/r - 2 \rho_0 m^2/r^2 + [m\rho_0/r - 3K\rho_0^{4/3}] \int_r^R r'^{-2} d(mr')]\}. (6.4)$$

An additional advantage of (6.4) as compared with (5.11) is that now all terms are of first order and the change (6.1) is not necessary. The integrals are straightforward using (4.1) and we get

$$E = 2 CM \rho_c^{-1/3} + F M^{7/3} \rho_c^{2/3} \text{ if } E \geq E_I, \quad F = F_1 (A/B) M^{-2/3} - F_2, \quad (6.5)$$

where $F_1 = 1.2483$, $F_2 = 0.3305$. We remark that $F_1 - F_2 = D_1 + D_2$ (see eq.(6.2)).

It may be realized that the energy as a function of $\rho_C$ has a minimum for any M. However for low M the minimum does not belong to the range of central densities $\rho_c$ allowed by the inequality $E \geq E_I$, which means that the equilibrium configuration possess local isotropy. Consequently, the standard theory applies in this range of masses (see comments at the end of section 4). For large M there are equilibrium configurations with local anisotropy, which we study in more detail in the following.

The transition from local isotropy to anisotropy happens for a mass quite close to the Chandrasekhar limit. (The transition mass corresponds to the simultaneous fullfilement of two conditions, namely that the central density $\rho_c$ minimizes E and that $E = E_I$). For any mass above this limit the equilibrium configuration of the star possess anisotropic pressure. The most interesting parameters in this case are the central density (obtained by minimization of eq.(6.5)) and the radius (obtained from eq.(4.1)). We get

$$\rho_c = C\, M^{-2/3}[F_1(A/B) - F_2 M^{2/3}]^{-1},\ R = 2.287\, C^{-1/3}\, M^{5/9}\, [F_1(A/B) - F_2\, M^{2/3}]^{1/3}. \quad (6.6)$$

We see that, for stars with the same A (i.e. the same chemical composition [4]) the central density decreases with increasing mass above the Chandrasekhar limit. The radius also increases but more slowly than the mass, so that the gravitational red shift at the surface, M/R, increases with mass.

## 7. DISCUSSION

I propose that equilibrium of a star requires that: 1) the general relativistic equation of hydrostatic equilibrium, eq.(5.3), is fulfilled, and 2) the total energy is a minimum, for a given number of baryons, with respect to two possible variations. The first is the spatial distribution of matter, given in spherical symmetry by the function $\rho(r)$. The second is the distribution of linear momentum at every point in the interior of the star, given by $x(r)$ in our approach. We remember that both the mass and the linear momentum contribute to gravity in the general theory of relativity. In the standard treatments the variations of the second kind are not allowed, that is local isotropy is imposed as an unjustified postulate.

I have shown that there are equilibrium configurations of white dwarfs above the Chandrasekhar limit. Furthermore the equilibrium corresponds to a minimum of the energy with respect to neighbour configurations, which will give stable equilibrium if we assume that in any slow oscillation of the star, eq.(5.3) holds approximately at all times. The question whether there is also stability against rapid oscillations should be investigated more thoroughly. The central density predicted for the high mass white dwarfs with local anisotropy is smaller than the central density of stars with mass close to the Chandrasekhar limit, but it should require a more detailed study the question whether that density is low enough to make the stars stable against neutronization. If the answer were in the affirmative there might exist stars with several solar masses sustained by the pressure of the degenerate electron gas. If these stars have become cold enough that they do not radiate, they could contribute to the missing mass in galaxies.

Our study involved essentially two approximations: post-Newtonian gravity and small local anisotropy. This has constrained us to deal with white dwarfs having masses close to the Chandrasekhar limit. For masses much higher a numerical solution of the exact equilibrium equation (5.3) would be necessary in order to know whether there are locally anisotropic equilibrium configurations.

The theory of equilibrium with local anisotropy here developped may be applied to other systems of astrophysical interest. Supermassive stars with local anisotropy are studied elsewhere [7]. A numerical study of neutron stars with local anisotropy would be also interesting. Finally I shall comment on the implications of the theory for main sequence stars, in particular the Sun.

The standard model of the Sun involves the "a priori" asssumption of isotropic pressure at every point [8]. But it might exist a model of the Sun with local anisotropy which is more stable than the standard model and still is compatible with all structure equations plus the observational data of mass, radius, luminosity, etc. If this is the case, the results of the present paper suggest that such a locally anisotropic model would have a smaller central density and, therefore, smaller central temperature. This would lead to a reduced production rate of $^7$Be and $^8$B neutrinos without changing the production of p-p neutrinos. In this way, allowing for a local anisotropy in the Sun might solve the solar neutrino problem.

## ACKNOWLEDGEMENT

I acknowledge financial support from DGICYT, Project No. PB-95-0594 (Spain).